\documentstyle[twoside,fleqn,espcrc2]{article}%
\def\tfig#1{{\xdef#1{Fig.\thinspace\the\figno}}
Fig.\thinspace\the\figno \global\advance\figno by1}
 \newcommand{\CZ}{{\cal Z}}
 \newcommand{\ptau  }{\partial_{\tau}}
 \newcommand{\CH}{{\cal H}}
  \newcommand{\CS}{{\cal S}}
\newcommand{\CF}{{\cal F}}
\newcommand{\CT}{{\cal T}}
 \newcommand{\CO}{{\cal O}}

 \newcommand{\CC}{ {\cal C}}
 \newcommand{\tr}{ {\rm  Tr}}

\newcommand{\p}{\partial}

 \newcommand{\CD}{{\cal D}}
\newcommand{\R}{\relax{\rm I\kern-.18em R}}
\font\cmss=cmss10 \font\cmsss=cmss10 at 7pt
\newcommand{\Z}{\relax\ifmmode\mathchoice
{\hbox{\cmss Z\kern-.4em Z}}{\hbox{\cmss Z\kern-.4em Z}}
{\lower.9pt\hbox{\cmsss Z\kern-.4em Z}}
{\lower1.2pt\hbox{\cmsss Z\kern-.4em Z}}\else{\cmss Z\kern-.4em Z}\fi}
\newcommand{\pl}{{\it  Phys. Lett.}}
 \newcommand{\cmp}{{\it Comm. Math. Phys. }}
\newcommand{\prl}{{\it  Phys. Rev. Lett.}}

\newcommand{\np}{{\it Nucl. Phys. }}

\newcommand{\CM}{{\cal M}}

\newcommand{\AmS}{{\protect\the\textfont2
  A\kern-.1667em\lower.5ex\hbox{M}\kern-.125emS}}

\title{Solvable statistical models on a random lattice}

\author{Ivan K. Kostov\address{Service de
 Physique Th\'eorique de Saclay,
 CE-Saclay, F-91191 Gif-sur-Yvette, France}
\thanks{On leave from the Institute for Nuclear Research and Nuclear
Energy, 72 Boulevard Tsarigradsko Chauss\'ee, 1784 Sofia,
Bulgaria}\thanks{{\tt kostov@amoco.saclay.cea.fr}}
         }

\begin{document}

\begin{abstract}

We give a sequence of equivalent formulations of the  $ADE$ and $\hat A\hat
D\hat E$ height models defined on a random triangulated surface: random
surfaces immersed in Dynkin diagrams, chains of coupled random matrices,
Coulomb gases, and multicomponent Bose and Fermi systems representing
soliton $\tau$-functions. We also formulate a set of loop-space  Feynman
rules allowing to calculate easily the partition function on a random surface
with arbitrary topology. The formalism allows to describe the critical
phenomena on a random surface in a unified fashion and gives a new meaning
to the $ADE$ classification.

\bigskip
{\sl
\centerline{Talk
delivered at the Conference on recent developments  in statistical
mechanics and quantum field theory}
\centerline{ (10 - 12 April 1995), Trieste, Italy.}}
\end{abstract}

\maketitle

\section{INTRODUCTION}

During  the  past decade,  stunning
progress has been made in understanding and solving
of statistical models on  two-dimensional
{\it random} lattices.
The interest in these systems arose mainly because of their interpretation
as models of discretized quantum gravity or, equivalently, ``noncritical"
string theories. The geometry of the lattice is considered here as an
additional   fluctuating variable.

The simplest statistical model on a random lattice is
the random lattice itself
(the target space consists of a single point).
This problem goes back to the problem of counting planar diagrams solved
in the seminal paper \cite{bipz} after
having been  reformulated in terms of a large $N$
matrix integral.  More recently, some nontrivial statistical
models have been solved on a random lattice following the
same principle: the Ising model \cite{bk}, the $O(n)$
model \cite{oon}, the $Q$-state Potts model \cite{potts},
and the $ADE$ height models \cite{Iade}. Each such system is equivalent to
an ensemble of coupled random matrices labeled by the points of the target
space of the model.
Besides the models listed above, there are
  many other solvable ensembles of matrix models
whose geometrical meaning still remains to be understood.

 The statistical  systems on random lattice  are interesting not only  as
toy models of discretized quantum gravity.
 In spite of their simplicity, the models on a random lattice still
contain  considerable information about the critical
behavior of the models on a regular lattice. Therefore they can be used
as an heuristic instrument
for studying complicated situations on regular lattices. For example, the
dilute critical regime of the $A  D  E$ models has been first discovered
first on a random \cite{Idis} and then on a regular lattice \cite{wn}.
Furthermore,  there are  infinitely many multicritical
 regimes  known on a random lattice, which are not yet considered on the
regular
lattice.
We believe that the correspondence between the models on random and regular
lattice merits to be understood on a deeper level and might help to
establish
  the  missing connections  between  the various resolution methods  used
by now.

 \smallskip

 In this talk  I would like to   review  the
 construction and the solution of the  $ADE$ and $\hat A \hat D \hat E$
height models on
a random lattice , as well as to present
 several  equivalent  formulations  of these models as simple integrable
systems.

 It is well known \cite{ciz}  that the  minimal two-dimensional
 conformal-invariant QFT  are classified by the
simply laced Lie algebras (i.e., these of the classical series $ A_{r},
D_{r}, E_{6},E_{7},E_{8}$).
  Each of these theories can be
constructed microscopically
as a lattice statistical model ( a $height$ model), whose local degrees of
freedom (heights) are labeled by
the points of  the Dynkin diagram of the corresponding Lie algebra
\cite{pas}.
 Similarly, the height models associated with the
extended  $\hat A \hat D \hat E$  Dynkin
 diagrams describe conformal invariant QFT with $C=1$ and discrete spectrum
of conformal dimensions.
The   height models in their dense (dilute) version can be mapped onto the
6-(19-)vertex model and solved using Bethe Anzatz \cite{abf}
(\cite{izk}).

The height models can be defined on an arbitrary surface made of triangles
or squares in such a way that the mapping onto a vertex model still holds.
In the next section  we will  remind the  definition of the
height models on an arbitrary   triangulated surface.  It requires
supplementary  Boltzmann factors  associated with the  points with nonzero
local curvature on the surface.
Then we describe briefly the loop gas representation and the loop diagram
technique that follows.

    The models on a random lattice become topological in the sense that
their correlation
functions do not depend any more on the distance.
The only parameters left are  the topology of the lattice, the volume and
the length of the boundaries and the boundary conditions.  The  local
scaling
operators  can be introduced by
shrinking a boundary with given boundary condition.

    In Sect. 3. the height models on random lattice are then reformulated
as systems of coupled random matrices, or,
in the eigenvalue representation, as Coulomb gas systems.

  A complete set of observables in the height models on a random  surface
is given by the set of all loop amplitudes or,
in  terms of free bosons or fermions, the   correlation functions
of currents.   A genus $g$, $n$-loop
amplitude is equal to the partition function of the model on a random
surface with the topology of a sphere with $g$ handles and $n$ boundaries
with given lengths.    In Sect. 4. we present  the loop diagram technique
established in \cite{hk} and
allowing to compute any loop amplitude on a surface with any topology.
 It will be formulated as a set of  Feynman rules involving
   propagators,
vertices (including tadpoles) of all topologies, and leg factors for the
external boundaries.
  A vertex of given topology  factorizes into a fusion
coefficient for the order parameters  of the height model and an
``intersection number" associated with the corresponding punctured
surface.
In Sect. 5. the partition functions   are  reformulated, using   the
vertex operator construction,
as  systems of free bosons or fermions (soliton $\tau$-functions).

 \section{INTEGRABLE HEIGHT MODELS LABELED
BY DYNKIN DIAGRAMS  }

\subsection{The target space }

   Let  $X$ be the   Dynkin graph a simply laced Lie algebra of rank $r$.
Such a graph   consists of a set of $r$ nodes, labeled by an integer
$height$ $x\in\{1,2,...,r\}$, and a number of bonds between nodes. Two
nodes are called adjacent ($\sim$) on $X$ if they are connected by  a
bond.
 The   graph  $X$ is defined by its   adjacency matrix
 \begin{equation}A^{xx'}=\cases{
1,& if $x\sim x'$  ;  \cr
0,& otherwise    \cr }
 \end{equation}
which is related to the Cartan matrix of the scalar products of the simple
roots
$\vec\alpha^1, ..., \vec\alpha^R$ by
\begin{equation}
\vec \alpha^x\cdot \vec \alpha^{x'} = 2\delta_{xx'} - A^{xx'}.
\label{cmt}
 \end{equation}

Each Dynkin graph $X$ can be extended to a  graph  $\hat X$ with $r+1$
nodes
  by adding an extra node representing the lowest root. The graph $\hat X$
can be considered as the  Dynkin graph of the corresponding affine Lie
algebra.
 For example, the  Dynkin diagram of the affine algebra $\hat A_{r}$
 $(R=1,2,...)$ represents a closed chain of $R+1$ points.
 One can consider as well the case $R=0$ where the target space,
  which will be denoted by $\hat A_0$, consists of one oriented link with
identified ends. The   connectivity matrix
of this target space is
$A_{00}=2$.

In what follows we
 assume that the target space $X$  is a Dynkin graph of an extended Dynkin
graph.

 The eigenvectors $S_x^{(m)}$ of the adjacency matrix are labeled by the
integer Coxeter exponents $m$
 \begin{equation}\sum_{x'} A^{xx'} S_{x'}^{(m)}=  2\cos {\pi m\over h} \ \
S^{(m)}_x.  \end{equation}
where $h$ is the dual Coxeter number of the Lie algebra.

 In order to
simplify the notations we shall always denote the eigenvector with largest
eigenvalue (the Perron-Frobenius vector)
by $S_x$,
\begin{equation}S_x =\cases{
S_x^{(1)} ,&  $ A,D,E$  ;\cr
S_{x}^{(0)},&  $\hat A,\hat D,\hat E$. \cr}
  \end{equation}
The normalized quantities
 \begin{equation}\chi _x^{(m)}={S_x^{(m)}\over S_x}.
 \end{equation}
  satisfy a closed (fusion)  algebra \cite{pas}
\begin{equation} \chi _x^{(m)} \chi _x^{(m')} =\sum _{(m'')}C^0 _{mm'm''}
\chi
_x^{(m'')} \end{equation}
 and  orthogonality conditions of the form
\begin{equation}\sum _x S_x^{2} \ \chi ^{(m)}_x\chi ^{(m')}
 _x= \delta _{m,m'} .  \end{equation}
More generally, we define the genus-$g$ fusion coefficients
  \begin{equation}
  C_{m_1\cdots m_n}^g = \sum_x    S_x^{2-2g}\
\chi^{(m_1)}_x\cdots\chi^{(m_n)}_x.
\label{fcns}
\end{equation}

\subsection{ Definition of the height models  }

 Let us remind the definition of the height model with target space $X$ on
 an arbitrary triangulated surface (two-dimensional
simplicial complex) $\CS $ \cite{Iade}.
  Each height configuration on $ \CS $ represents a
   map $\CS \rightarrow X$ such that the heights
 $x(\sigma),x(\sigma') $ of any two adjacent sites
 $\sigma,\sigma'$ of $\CS $ are adjacent or equal in the target space.
 The partition function of the height model  defined on the surface $\CS$
is
equal to the sum over all
 height configurations $\CS \rightarrow X$
 \begin{equation}
 	\CF[\CS ]=\sum_{\CS \rightarrow X}
 	W(\CS \rightarrow X).
 	\label{pfxs}
 \end{equation}
  with   Dirichlet boundary conditions , i.e., constant height $x_i$
along the connected components $\CC_i$ of the boundary $\p \CS$, imposed.
 The Boltzmann weight of each height configuration is a product
 of factors associated with the vertices $\sigma$ and the triangles
 $\triangle$ of $\CS $
 \begin{equation}
 	W(\CS \rightarrow X)=\prod_{\sigma\in \CS }
 	W_{\bullet }[x(\sigma)]\prod_{\triangle\in \CS } W_{\triangle}
 	[x(\triangle)].
 	\label{bwpt}
 \end{equation}
 Note that in  the case  of  a  regular  triangular   lattice,  where
exactly 6 triangles
 meet  at     each
  vertex, the  Boltzmann weights associated with
   the vertices can be distributed among the adjacent triangles.
   In the case of a lattice with defects we have to take into account the
     local curvature and the the form (\ref{bwpt}) of the Boltzmann
weights is
   more appropriate. The local Boltzmann weights  are expressed through the
   components of the Perron-Frobenius vector $S_x$ as follows
\begin{eqnarray}
W_{\bullet }(x)=S_x\\
   W_{\triangle} (x_1,x_2,x_3)=
  {\delta_{x_1x_2}A_{x_2x_3}A_{x_3x_1}  \over\sqrt{S_{x_1}}}
\nonumber \\
 + {\delta_{x_2x_3}A_{x_3x_1}A_{x_1x_2}  \over\sqrt{S_{x_2}}}+
{\delta_{x_3x_1}A_{x_1x_2}A_{x_2x_3}
 \over\sqrt{S_{x_3}}}.
 \label{tttr}
\end{eqnarray}

\subsection{The height models on a random surface}

 The height model on a random lattice  is the statistical ensemble
of all height configurations  $\CS\to X$ where the
 surface $\CS$ is treated as
 as an additional  dynamical variable.
   The partition function of the model is defined as the sum over all
surfaces $\CS$ and all maps $\CS\to X$ satisfying certain conditions.
Thus the height model on a random lattice can be also viewed as a problem
of
a random surface $\CS$ immersed in the discrete target space $X$.

 Denote by   $\Sigma ^{g,A}_{l_1,...,l_n}$
 the ensemble of all triangulated  surfaces $\CS $ with genus $g$, area
$A$,
 and  with   boundary $\partial \CS $
 consisting of $n$ loops $C_1,...,C_n$ with lengths $l_1,...,l_n$
For each surface $\CS $ we define the  partition function $\CF[\CS ,
x_1,...,x_n]$
where $x_i$ denotes the height of  loop $C_i$.   The genus
 $g$, $n$-loop amplitude reads
   \begin{eqnarray}
 	\CF^{g,A}_{l_1,x_1;..., l_n,x_n }=\sum _{\CS}\CF(\CS ; x_1,...,x_n),
\nonumber \\
 \CS \in \Sigma ^{g,A}_{l_1,...,l_n}.
 \label{mcnl}
 \end{eqnarray}

 The  partition function of the height model with target space $X$ is by
definition the exponent
\begin{equation}
\CZ [\kappa, T  ,  g_{kx}]=e^{\CF[\kappa, T,    g_{kx}]}
\label{eexx}
\end{equation}
of the
generating function for the  loop amplitudes
 \begin{eqnarray}
 	\CF[\kappa, T,   g_{k,x}]= \sum_{g=0}^{\infty} \kappa ^{ 2g-2+n}\sum
_{A\ge 0} {1\over T^A}\sum_{n=0}^{\infty}{1\over n!}\nonumber \\
   \sum_{\l_i > 0}\sum_{ x_i\in X }
   {{g_{l_1x_1}\over  S_{x_1}^{l_1/2}}...{g_{l_nx_n}\over
     S_{x_n}^{l_n/2}}}
    \CF^{g,A}_{l_1,x_1;...,l_n,x_n }.
  \label{cnpf}
 \end{eqnarray}

The mapping of the  height models  onto a gas of loops yields a powerful
combinatorial method
allowing to write a complete set of
equations for the loop amplitudes.
Let us sketch the loop gas method.
 For given triangulated surface $\CS $,
   each height configuration defines a set of nonintersecting loops
   on the dual tri-coordinated graph.
 For example, if $\CS $ is
     the regular triangular lattice, the  loops live on the  dual honeycomb
      lattice.

The loops represent the domain walls
   separating the domains of constant height $x$. The Boltzmann
   weight of a loop configuration can be organized as  a product of
factors
   associated with  its connected domains and the boundaries of the
surface \cite{Iade}.
  A domain of height $x$ contributes a factor   $S_x^{2-2g-n}$, where $n$
is the number
   of its boundaries and $g$ is the enclosed genus. With the definition
(\ref{cnpf}) of the generating function, this rule   applies also to the
domains
adjacent to the boundaries.   The Boltzmann factors associated with a
boundary of
length $l$ and height $x$ is $g_{lx}$.

   The fact that the Boltzmann weights of the domains depend only on
   their topology allows   to perform the
   sum over loops  and the sum over surfaces
    in the opposite order. Namely, we first   perform the sum
    over the shapes of the domains   and
    afterwards the sum over the topologically inequivalent loop
configurations.
    The result of the summation can be expressed as a sum of
    connected graphs whose vertices and lines label correspondingly the
domains and the
    domain walls.
     Let $W^{(g)}_{l_1,...,l_n}$ be the number of
   of  domains
     of genus $g$   and $n$ connected boundaries of lengths
   $l_1,..., l_n$ (we assume that each boundary has a marked point).
   Then the  partition function  (\ref{cnpf})
   is equal to the sum of all connected
   Feynman graphs composed from  the propagator
   \begin{equation}
   G_{l,x;l',x'}=A_{x,x'}\ {T^{-l-l' }\over l+l'}  {(l+l')!\over l!\ \
l'!},
   	\label{prp}
   \end{equation}
   the vertices
   \begin{equation}
   V^{(g)}_{x;l_1,...,l_n}= W^{(g)}_{l_1,...,l_n}\Bigg({\kappa\over
S_x}\Bigg)^{ 2g-2+n},   (n\ge 2)
    	\label{stvtx}
    \end{equation}
 and  the tadpoles
   \begin{equation}
   V^{(g)}_{x,l}=W^{(g)}_{l_1}\Bigg({\kappa\over S_x}\Bigg)^{ 2g-1}	+
\delta _{g, 0} \ g_{lx}.
   	\label{tdpl}
   \end{equation}

The large $l$ asymptotics of the quantities  $W^{(g)}_{l_1,...,l_n}$  is
given by   the loop amplitudes for the
so called "topological gravity"  whose explicit form is known.
 The corresponding loop Feynman rules are derived in Section 5.

 The loop gas representation is encoded in  the following
  Ward identities for the partition function \cite{Iade}
 (\ref{eexx})
  \begin{eqnarray}
  	 \Big[ -{\p\over \partial  g_{nx}} +\sum _{k\ge 3}k g_{kx} {\p\over
\partial  g_{ n-2-k,x} }
  +\sum_{k=0}^{n-2}
  	{\p\over \partial  g_{kx}}\nonumber \\ {\p\over \partial  g_{n-2-k,x} }
  	+\sum_{x'}A^{xx'} \sum_{k,k'\ge 0}
{(k+k')!\over k!\ \ k'!}
 {1\over  T^{k+k'+1} }\nonumber \\
{\p\over \partial  g_{ n-1+k,x}} {\p\over \partial  g_ {  k'x'}} \Big]
\CZ =0\ \ (x\in X, n=1,2,...)
  	\label{lebr}
  \end{eqnarray}

The coefficients of the genus expansion of (\ref{lebr})
give identities relating the loop amplitudes of different genus.
 The leading term  in the expansion in $\kappa$ leads to  a closed
equation for
 the disc
amplitude $W_{x,l}= \p \log \CZ  /  \p g_{k,x} |_{\kappa\to 0}$
  \begin{eqnarray}
  	 \Big[ W{n,x} =\sum _{k\ge 3}k g_{k,x} W_{ n-2-k,x}
  +\sum_{k=0}^{n-2}
  	W_{k,x}\nonumber \\ W_{n-2-k,x}
  	+\sum_{x'}A^{xx'} \sum_{k,k'\ge 0}
{(k+k')!\over k!\ \ k'!}
 {1\over  T^{k+k'+1} }\nonumber \\
W_{ n-1+k,x} W_ {  k',x'} \Big]
 \ \ \  (x\in X, n=1,2,...).
  	\label{plln}
  \end{eqnarray}

   \section{  ENSEMBLES  OF COUPLED RANDOM MATRICES}

   The crucial observation allowing to establish the equivalence with
ensembles of random matrices
  is that the   the vertices (\ref{stvtx}) can be represented as the  loop
correlators in a
Gaussian   $N\times N$ matrix model with
$N=S_x/\kappa$
   \begin{eqnarray}
   \sum _{g=0}^{\infty} N^{2-2g-n} W^{(g)}_{l_1,...,l_n}
     	 \nonumber \\
=\int D{\bf  M}  e^{  - \frac{N}{2}\tr {\bf M}_x^2 }	\  \tr  { \bf
M}^{l_1}... \tr{ \bf  M}^{l_n}.
   \label{gint}
   \end{eqnarray}
 Therefore, if we  associate with each
height $x$ a hermitian  matrix
variable ${\bf M_x} $ of size $N_x\times N_x$
where
\begin{equation}
  	\ \ \ \ \ \ \
\ \ \ \ \ \ \ \ \ \ \ \ N_x={S_x\over\kappa},
  	\label{nnx}
 \end{equation}
     the
perturbative expansion of the
matrix integral
\begin{eqnarray}
\CZ_X[g_{lx}, N_x; x\in X] = \int \prod_{x}  D{\bf M}_x\nonumber \\
\exp\Big[\tr \sum_{x\in X}
\Big( - {N_x\over 2}
 {\bf M}_x^2
 +\sum_{l=1}^{\infty} g_{lx}{\bf M}_x^l\Big)\nonumber \\
 +\frac{1}{2} \sum_{x ,x'}\sum_{l,l'}\tr {\bf M}_x^l  G_{lx;l'x'}\tr {\bf
M}_{x'}^{l'}
\Big]
\label{mmod}
\end{eqnarray}
around the gaussian measure will give the partition function  $\CZ= e^\CF$
 (\ref{cnpf}) \cite{Imat}.     One can check that the Ward identities
(\ref{lebr}) follow from the
   translational invariance of the integration measure in (\ref{mmod}).

The sum in the exponent with the coefficients (\ref{prp}) is a
simple logarithm and after  shifting the matrix variables
as
\begin{equation}
  {\bf M}_x\to {\bf M}_x+T/2
  \label{SHI}
  \end{equation}
we write the
  integral (\ref{mmod})
as
\begin{eqnarray}
\CZ_X[ V^x; N_x]= \int \prod_{x\in X}  D{  \bf  M}_x  e^{
-\tr V^x({\bf M_x} )}\nonumber\\
  \ \ \ \ \ \ \ \ \prod_{x,x'}
 \Big| \det   (   1\otimes {\bf M}_x+
{\bf M}_{x'} \otimes 1 ) \Big| ^{-A^{x,x'}/2 }
\label{mpatr}
\end{eqnarray}
 where  the coefficients in the expansion
   of the potential
 \begin{equation}
	V^x(z)= \sum_n T_{n}^x z^n
	\label{popo}
\end{equation}

In particular, the $\hat A_0$ model is
 dual  to the $O(2)$ model on a random lattice
\cite{oon}
and its partition function is given by the following $N\times N$ matrix
integral
\begin{eqnarray}
\CZ_{\hat A_0}[V,N]= \int  D{\bf M}{ e^{-\tr V({\bf M})}
 \over \Big| \det   (   1\otimes {\bf M}+
{\bf M}\otimes 1 ) \Big|  }.
\label{mpotr}
 \end{eqnarray}

   \subsection{Coulomb gas representation}

   The  integrand in  (\ref{mpatr}) depends only on the
eigenvalues $\lambda_{ix}, i=1,2,...,N_x,$ of the   matrix
variable   ${\bf M_x} $. Therefore we can  retain only the radial part
of the integration measure $d{\bf M}_x \sim \prod_{i} d\lambda_{ix} \prod_{i<j}
(\lambda_{ix}-\lambda_{jx})^2$  and write the partition function
(\ref{mpatr}) as
  \begin{eqnarray}
\CZ_{X}[ V^x, N_x; x\in X] =     \int  \prod _{ x\in X} \prod_{i=1}^{N_x}
d\lambda_{ i,x} \nonumber \\
  e^{ -V^x ( \lambda_ {i,x}) }
\prod_{x,x'\in X}
{\prod_{i\ne j} (\lambda_{ix }-\lambda_{j x'})^{\delta_{x,x'} }\over
   \prod_{i,j}  |\lambda_ {ia} +\lambda_ {ja'}|^{ A^{xx'}/2}}.
\label{mtria}
\end{eqnarray}

 Similar integrals have been introduced by
the ITEP group as ``conformal matrix models'' \cite{cmm}.
 To make the connection with the construction in \cite{cmm}
 let us divide  the nodes of the Dynkin diagram $X$ into two
groups
   (say, black and white) so that the nearest neighbors of each node
   have the opposite color and change the sign of the $\lambda$-variables
   associated with the black points.
Denote by $\varepsilon_x$ the color (white ($+$)
or black ($-$)) of the point
$x\in X$.  Then the integral can be
 written in the form
 \begin{eqnarray}
  \CZ_X[  V^x, N_x; x\in X ] =     \int \prod _{ x\in X} \prod_{i=1}^{N_x}
     d\lambda_{ i,x}  \nonumber \\
 e^{ -V^x (\varepsilon_x \lambda_
{ix}) }  \prod_{(i,x)\ne (j,x')}
  (\lambda_{ix }-\lambda_{j x'})^{ \frac{1}{2}  \vec\alpha^x\cdot
   \vec \alpha^{x'}}
\label{defev}
\end{eqnarray}
where we used the definition of the Cartan matrix (\ref{cmt}).
This integral can be interpreted as
  the partition function of $N=\sum_{x} N_x$ Coulomb particles with
vector charges proportional to the simple roots of the Lie algebra  with
Dynkin diagram $X$, restricted on a line  and interacting with  a common
potential.

The integral over the eigenvalues  (the positions of the charges)
is well defined in the large $N$ limit only if the
potential  keeps  the white charges in the  negative half-space $\lambda
<0$  and the black charges  in the positive half-space $\lambda >0$.
Then a  nonsingular  large $N$ limit exists since the charges of the
same color do not interact and the
the black and white charges that attract each other
are separated by a potential wall.
The critical situation arises when the   edges  of the black and white
charge
densities meet  at  the origin. The
critical singularity is explained in terms of the original statistical
model by the dominance of
triangulated surfaces with infinite area.

  In what follows we prefer to  keep  the representation (\ref{mtria})
of the Coulomb system, which is more naturally related to our original problem,
and which can be considered as a generalization of the
partition function of the $\hat A_0$ model
  \begin{eqnarray}\CZ_{\hat A_0} [V,N] =   \int \prod _{i=1}^N d\lambda_i \
e^{-V(\lambda_i)}  {\prod_{i< j} (\lambda_i-\lambda_j)^2\over
\prod_{ij}|\lambda_i+\lambda_j|}.
\label{pjiao}
\end{eqnarray}
 The existence of a  nonsingular thermodynamical limit then implies that the
 distribution of the eigenvalues
 has its support along the   negative real axis   for all $x$.

    Therefore   the integration over
the eigenvalues $\lambda_{ix}$ can be  restricted from the very beginning
to the interval $\lambda <0$. This restriction will modify the nonperturbative
sector but the genus expansion of the partition function will remain the same.

  Note also that, since the point $\lambda =0$ is outside the support of the
spectral
density, one can consider a more general potential containing negative powers
of $\lambda$
as well. On the other hand, the algebra simplifies considerably if the
even powers in the expansion of the potential $V(\lambda)$ are suppressed.

 It is sometimes more advantageous to consider instead of (\ref{mtria}) the
$canonical$
 partition function
 \begin{equation}
 \CZ_X[ V^x, \mu^x]=\sum_{ N_x\ge 0}\prod_x { e^{  \mu^x   N_x}\over \prod
_{x}N_x!}
  \ \CZ_X[ V^x,   N_x].
 \label{canno}
 \end{equation}
 We will assume in the following that
the chemical potential $  \mu^x $  for the charges of type  $x$
 is absorbed in the potential  $V^x$ as a constant term.

   \subsection{The partition function of
the $\hat A_{r}$ model as a
Fredholm determinant}

Let us introduce  the integration kernels \begin{eqnarray}
K_{x,x'}(\lambda, \lambda') = {e^{-\frac{1}{2} [  V^x(\lambda)+
V^{x'}(\lambda')]}\over \lambda +\lambda'} \nonumber \\
(x=0,1,...,R; \ \ x'=x+1)
\label{kernnn}
\end{eqnarray}
where the integration goes along the negative real axis.
 With the help of the Cauchy identity
\begin{equation}
{\prod_{i<j}(\lambda_i-\lambda_j)(\mu_i-\mu_j)\over
\prod_{i,j} (\lambda_i+\mu_j)}= \det_{ij} {1\over \lambda_i+\mu_j}
\label{Cauch}
\end{equation}
we can write  the canonical partition function of the $\hat A_r$ model (whose
target
space $X$ is a circle of $R+1$ points) as   a Fredholm determinant
\begin{eqnarray}
\CZ_{\hat A_r} =    Det
(1+K_{01}K_{12}K_{23}...K_{R0}).
\label{frdh}
\end{eqnarray}

In particular, the canonical partition function for the $\hat A_0$ model  is
given by
\begin{eqnarray}
\CZ_{\hat A_0} = \sum_{N=0}^{\infty} {1\over N!} \prod_{i=1}^N d\lambda_i
	  \det_{ij} K(\lambda_i,\lambda_j)\nonumber\\ =  {\rm Det}(1+K)
\label{odva}
\end{eqnarray}
 where
\begin{eqnarray}
K(\lambda, \lambda') = {e^{-\frac{1}{2} [  V(\lambda)+
V(\lambda')]}\over \lambda +\lambda'} \nonumber \\
V(\lambda)= -\mu  -\sum_{n}  t_{2n+1}\lambda^{2n+1}.
 \label{kernoo}
\end{eqnarray}

   Fredholm determinants like   (\ref{odva})
give the correlation functions in various statistical or QFT models.
The simplest case is the two-spin correlation function in the Ising model
\cite{mctw} satisfying the  Painlev\'e III equation. More recently, this
Fredholm determinant has been considered
in the context  of self-avoiding polymers on a cylinder
\cite{saz}.

The representation (\ref{frdh}) of
  the partition functions of the $\hat A_r$ models
leads   to the problem of diagonalization of the
 kernel (\ref{kernnn}).  This problem  has been solved  explicitly only in
the  case of a potential
 $V(z)= t_1\lambda+t_{-1} \lambda^{-1}$ \cite{mct}.
  An original  method to find the eigenfunctions in this case has been
suggested by M. Staudacher \cite{Matth}. Unfortunately the method does not work
more complicated potentials.
 The idea is to look for
 a differential operator
commuting with the  integral  operator represented by the kernel
$K$ and therefore having the same spectrum of eigenstates.
If we introduce the parametruzation $\lambda = -e^{u}\
(-\infty<u<\infty)$,
then the    kernel (\ref{kernnn}) becomes
\begin{eqnarray}
K(u,u')= {e^{ - \frac{1}{2}[v(u)+v(u')] }\over \cosh {u-u'\over 2}}
 \label{trep}
\end{eqnarray}
with potential
\begin{equation}
 v(u) \equiv V(e^u) = -\mu -t _1e^u -t_{-1} e^{-u}
 \end{equation}
 and  nonrestricted homogeneous measure $du$.
It is easy to check that the linear operator with kernel
(\ref{trep}) commutes with the differential operator
\begin{equation}
  H= {\partial^2 \over \partial u^2} -   [\frac{1}{2} v'(u)]^2
\label{oiuyuy}
\end{equation}
 and therefore has the same set of eigenstates.

    \subsection{ Loop equations  the $ADE$ and $\hat A\hat D\hat E$ matrix
models}

 The resolvent
 \begin{eqnarray}
W_x(z)= \lim_{N\to \infty} \langle   \sum_{i=1}^{N_x} {1\over
z-\lambda_{ix}}\rangle
\label{ohh}
   \end{eqnarray}
 represents a meromorphic function of $z$ with a
cut along the
support $[a,b] \ (a<b<0)$ of the classical eigenvalue density
 and behaves at infinity as
 \begin{eqnarray}
W_x(z)= {N_x\over z}+\CO({1\over z^2}).
\end{eqnarray}
The loop equations (\ref{lebr})   can be
derived in the matrix model from the translational invariance of
the matrix integration measure.
The classical ($\kappa \to 0$) loop equation for the  resolvent $W_x(z)$ reads
 \cite{Idis}
 \begin{eqnarray}
W_x(z)^2+\oint_{\CC_-}{dw\over 2\pi i}{ 1\over z-w}W_x(w) [-\partial
V^x(w) \nonumber \\
 + \sum_{x'}A^{xx'} W_{x'}(-w)]=0\label{cllpe}
\end{eqnarray}
where the contour
 $\CC_-$ encloses
the point $z$ and the cut $[a,b]$ of $W_x(z)$
and leaves outside the cut $[-b,-a]$ of
$W_x(-z)$.

 The exact loop equations are obtained from (\ref{cllpe}) by replacing
$W_x(z)$ with the
  loop insertion operator
 \begin{eqnarray}
{d\over dV^x(z)}= \sum_{n=1}^{\infty} z^{-n-1}
{\p\over \partial  T_{n }^x}
\label{lliE}
\end{eqnarray}
 and represent  the
following quadratic differential constraints on the partition
function
 \begin{eqnarray}
  \oint_{\CC_-}{dw\over 2\pi i}{ 1\over z-w}
\CT^{(x)}(w) \ \ \CZ_X =0, \ \ \ x\in X \ \ \
\label{vviir}
\end{eqnarray}
where
 \begin{eqnarray}
 \CT^{(x)}(z) = {d^2\over dV^x(z)^2}+
   \Big(-\partial  V^x(z)\nonumber \\
 + \sum_{x'}A^{xx'} {d\over dV^{x'}(-z)}\Big)
      {d\over dV^x(z)} .\label{qqst}
\end{eqnarray}

Let us introduce  the current-like field
\begin{eqnarray}
 \partial \hat\Phi_x(z) = {d\over dV^x(z)}-  \partial V_x(z).
\label{ocon}
\end{eqnarray}
where
  the covariant components $V_x(z)$ of the potential are defined  by
 \begin{eqnarray}
V^x(z)=2V_x(z)-\sum_{x'}A^{xx'}V_{x'}(-z).
\label{cocon}
\end{eqnarray}
  Then the operators (\ref{qqst}) can be  written in the form
  \begin{eqnarray}
   \CT^{(x)}(z)=\partial \hat\Phi_x(z) [\partial \hat\Phi_x(z)+
   \sum_{x'} A^{xx'} \partial \hat\Phi_{x'}(-z) ].
\label{oconn}
\end{eqnarray}
  The classical   expectation value $\p \Phi_x(z)$  of the current
 (\ref{ocon})  is  related to the
resolvent
(\ref{ohh}) by
  $ \partial \Phi_x(z) = W_x(z)- \partial V_x(z) $ and
 is completely determined by the conditions
 \begin{eqnarray}
\Phi_x(\lambda+i0) + \Phi_x(\lambda-i0) -\sum_{x'}A^{xx'}\Phi_{x'}(-z)
\nonumber \\
=0,  \ \ \ \ (a<\lambda<b);
 \\
\Phi_x(z)= -V_x(z)+ N_x\ln z+\CO({1\over z}),   (z\to \infty),
\label{asil}
\end{eqnarray}
which are obtained by taking the imaginary part of
the classical loop equation (\ref{cllpe}).
The first of these equations  means that the meromorphic function
 \begin{eqnarray}
 \hat \Phi^x(z)=2\hat \Phi_x(z)-\sum_{x'}A^{xx'}\hat \Phi_{x'}(-z).
\label{co ocon}
\end{eqnarray}
has vanishing real part along the cut $[a,b]$.

The constraints  (\ref{vviir})  form  a set of mutually
commuting Virasoro algebras,  one for each
point $x$ of the target space.
One can interpret the Virasoro constraints for given point $x$ as
the loop equations for the topological gravity \cite{dvv}, with potential
depending on the
fields associated with the adjacent points $x'$.
In this way the solution of the topological gravity
  appears as the mean field problem in our models.
In the next section we formulate a  collective  field approach utilising the
solution of the topological gravity.

\section{LOOP DIAGRAM TECHNIQUE}

The tree (genus zero) amplitudes can be calculated with the
effective collective field theory in which the eigenvalues
of the random matrices are considered as
a continuous classical liquid.
The quantum fluctuations of the liquid interact in a complicated way
but the interaction potential can be determined exactly.
The strategy is  based on the exact solution of the mean field problem,
namely the integral over the positions $\lambda_{ix} \ (i=1,...,N_x)$ of
the charges of type $x$ for fixed distribution of the other charges.
This leads us to a  one-matrix model for generic potential
whose solution is known. The free energy of this effective
one-matrix model gives the interactions of the vibration modes
of the eigenvalue liquid.

 Let us introduce collective field $  \Phi_x(z)$   and the corresponding
lagrange multiplier field $\tilde V^x(z)$
    by   inserting the
identity
\begin{eqnarray}
1= \int \CD\Phi\CD \tilde V \prod_x  \ \exp \Big( {1\over 2\pi i} \oint
_{\CC_-}
 d[ \Phi_x(z) \nonumber \\ -V_x(z)
-\sum _{i=1}^{N_x}
\ln (z-\lambda_{ix})]
 \tilde V^x(z) \Big)
\label{jacc}
\end{eqnarray}
in the  r.h.s. of (\ref{mtria}).   The integration with respect to the
$\lambda_{ix}, i=1,... N_x,$ yields the    mean field free energy
$F[\tilde V^x]$  defined by
 \begin{eqnarray} e^{ F[V]}= \int  \prod_{i=1}^N   d\lambda_{i } e^{-
V(\lambda_i)} \prod_{i<j}(\lambda_{i }-\lambda_{j })^2.
\label{fren}
\end{eqnarray}
 After  this change of variables the
  original partition  function
 becomes
  a functional integral over the collective fields $\Phi$ and $\tilde V$
with
nonrestricted homogeneous measure
\begin{eqnarray} \CZ=\int   \CD  \Phi  \CD \tilde V \
 e^{  - \CS  [\Phi_x , \tilde V^x;x\in X] } ,\\
\label{inntt}
    \CS  =  \sum_x \Big[-F   [\tilde V^x] - \oint {d
\Phi_{x'}(z) \over 2\pi i}
\tilde V^x(z)\Big]   \nonumber \\
    +\frac{1}{2} \sum_{x,x' }A_{xx'}   \oint {d \Phi_{x'}(z') \over 2\pi i}
\oint { d \Phi_x(z) \over 2\pi i}
 \ln ( z+ z')     .
 \label{acctt}
\end{eqnarray}
 The dependence on the potential $V^x(z)$ is through the
asymptotics of the field $\Phi_x$ at $z\to \infty$.

 The string propagator and vertices  are obtained by expanding the
effective
action  (\ref{acctt}) around the
 mean field $\Phi_{x}^{c},    \tilde V^{x}_c$ determined by the large $N$
saddle point equations.
The dependence on the string coupling constant $\kappa = S_x/N_x$ is
through   the genus expansion of  the
 effective potential
 $F[\tilde V^x]= \sum _{g\ge 0}  N_x^{2-2g}\CF^{(g)}[\tilde V^x] $.
It is convenient to make one more change of variables and introduce  the
field $\tilde \Phi_x(z)$  defined by
\begin{equation}
\partial  \tilde \Phi_x(z)= -{dF^{(0)}[\tilde V^x]\over d \tilde V^x(z)
}-\frac{1}{2}\partial  \tilde V^x(z) \end{equation}
  as an independent fluctuating variable.
The  fields
$\tilde \Phi_x(z)$ and $ \Phi_x(z)$ have the same
classical values  $ \Phi_x^c(z)$ but on the quantum level they play
different roles  : the external source is coupled  only to the field   $
\Phi_x(z)$
  while the interactions affect only of the field
 $\tilde \Phi_x(z)$.

It is of course possible to  consider the discontinuities
  \begin{equation}
\phi_x (\lambda) = -{1\over \pi} \Im
\Phi_x(\lambda),\ \ \ \tilde
\phi_x (\lambda) = -{1\over \pi} \Im
\tilde \Phi_x(\lambda)
\end{equation}
  as independent functional variables.
  In any case, the functional integration measure in (\ref{inntt})
is  defined only after introducing a mode expansion for the two   fields.
 \subsection{Saddle point}

    Upon an appropriate rescaling   of the $\lambda$-variable,   the
support $[a,b]$ of  the classical spectral density
 $\rho^c(\lambda) = - {1\over \pi} d \Im\Phi_x^c(\lambda) /d\lambda$
can be taken to be  the interval  $[- L, -1]$ on the negative real axis.

The solution of the saddle point  equation in the continuum limit
$L\to\infty$ reads,
 up to an arbitrary normalization,
\begin{eqnarray}
X=ADE: \ \ \ \ \ \ \  {d \over dz}\Phi_x^c(z) =\nonumber\\
- {  S_x \over \kappa} \ { (z +\sqrt{z^2-1})^{\beta} +(z -\sqrt{z
^2-1})^{\beta} \over
\  2\pi  {\beta}  |\sin \pi {\beta} | }
\label{reshe}
\end{eqnarray}
 where
\begin{equation}
\beta=1\pm {1\over h} +2m
\label{dfyu}
\end{equation}
and  the string coupling constant scales as $\kappa  \sim
L^{1+{\beta}}$.    The different  branches  $(\pm, m)$ correspond the
different
critical regimes of the
height model (for details see \cite{Idis}  and \cite{Imatth}).  The
regimes  $(-,0)$ and  $(+,0)$  are sometimes referred as $dense$ and
$dilute$ phase
of the height model.
   The central
charge of the  corresponding
conformal field theory is
\begin{equation} C= 1-6 {({\beta} -1)^2\over {\beta}} .
\end{equation}
  The saddle-point solution for the case $X=\hat A\hat D\hat E$ is
obtained as the limit $\beta \to 1$ of
(\ref{reshe}).

 Our aim now is   to
  define the Hilbert space of one-loop states,  choose a complete
orthonormalized  set of eigenstates of the  quadratic action, and finally
express
the interactions in terms of the mode expansion of the collective fields.

\subsection{Gaussian fluctuations  }

It is consistent with the  perturbative expansion to assume that the
fluctuating fields are again
  supported by a semi-infinite  interval, but its right end can be
slightly displaced with respect to its
saddle point value $  -1$    due to  fluctuations.

Inserting the
genus-zero contribution to the mean field free energy  (\ref{fren})
\begin{equation}
F^{(0)}[\tilde V_x]=
 -{1\over \pi} \int  \Re \tilde\Phi_x
  d  \Im \tilde\Phi_x
\end{equation}
  we find for  the tree-level  action
\begin{eqnarray}
 \CS ^{(0)} = {1\over \pi} \sum_x
\int  \Big\{   \Re \tilde\Phi_x (\lambda)  d   \Im \tilde\Phi_x (\lambda)
\nonumber \\
+\big[  \sum_{x'} A^{xx'} \Phi_{x'} (-\lambda)
 - 2\Re \tilde\Phi_x (\lambda)  \big]   d \Im \Phi_x (\lambda) \Big\}.
\label{ccii}
\end{eqnarray}

The   tree-level  effective action (\ref{ccii})
 can be split into   gaussian and interacting parts
\begin{equation}
  \CS ^{(0)} =\CS^{\rm free}  + \sum _{n\ge 3} \CS ^{(0,n)}
\label{ssss}
\end{equation}
 where  $ \CS ^{(0,n)}$ is the genus-zero  $n$-loop   interaction.

 The gaussian fluctuations of the collective fields are those that do not
shift the edge of
 the eigenvalue interval; the  fluctuations    displacing  of the edge
are taken into account by the
 nongaussian terms in the effective action  that  represent  the  tree
level  $n$-loop interactions.

We therefore define the  Hilbert space $\CH$
as the space of real functions     defined on $X\times [-\infty, -1]  $
or, equivalently,
the space of analytic fields having a cut along the interval $[-\infty,
-1]$.

For the purpose of diagonalizing the quadratic action it is very useful
to   introduce
    the map
     \begin{equation}
\ \ \ \ \ \ \ \ \ \ \ \ \ \ \ \ \ z(\tau)   =  \cosh \tau   \label{prms}
\end{equation}
    transforming  the space of
meromorphic functions  in  the   $z$-plane  cut along the interval
 $[-\infty , -1]$ into  the space of entire even analytic functions of
$\tau$.
In the following we will keep the same letters for the analytic fields
considered as functions of $\tau$ and denote  $ \Phi(\tau) \equiv  \Phi(z(
\tau)) $.
 The  disc amplitude  (\ref{reshe})   as a function of   $\tau$  reads
 \begin{equation}  {d  \over dz}   \Phi_x^c (\tau)  =  {S_x \over
\kappa}\ \
{\cosh \beta \tau \   \over 2\pi \beta  |\sin \pi  \beta |   } .
\end{equation}

The operations $\Im$ and $\Re$ in the $z$-space   become finite-difference
operators in the  $\tau$-space. Denoting $ \partial   = \partial/\partial
\tau) $,we can write
 \begin{eqnarray}
 \Im \Phi(-z)= \sin \pi \partial   \Phi(\tau),
\nonumber \\
 \Re \Phi(-z)= \cos \pi  \partial   \Phi(\tau)
\label{imre}
\end{eqnarray}
 It is   clear that
the   plane waves
\begin{equation}
\langle x, \tau |m, E \rangle = S^{(m)}_x  e^{iE\tau}
\label{fint}
\end{equation}
 form a
 complete set of (delta-function) normalized wave functions diagonalizing
the quadratic action
  The latter is  given by the expression  (\ref{ccii}) where the
integration is restricted
to the interval $\lambda \le 1$.
 According to (\ref{imre}), the Fourier components of the fields
$\tilde\phi$ and $\phi$ are related to these of $\tilde\Phi$ and $\Phi$ by
$\tilde\phi_{(m)}(E)= {1\over\pi}\sinh (\pi E ) \tilde\Phi_{(m)}(E), \
\phi_{(m)}(E)= {1\over\pi}\sinh (\pi E ) \Phi_{(m)}(E)$.
The action  (\ref{ccii}) reads, in terms of these fields,
\begin{eqnarray}
  \CS^{\rm free}[\phi,  \tilde\phi]=\sum _{p=m/h}  \int
_{0}^{\infty}{dE\over 2\pi}\Big[
\phi \ { \pi E\cos\pi p   \over  \sinh \pi E}\phi
   \nonumber \\
  +  ( \tilde\phi- 2 \phi    ) \ { \pi E\cosh\pi E  \over  \sinh \pi E}
\ \tilde\phi
     \Big].
\label{elisav}
\end{eqnarray}
  By inverting the quadratic form in (\ref{elisav}) we find the
  propagators  in the $(E, p={m\over h})$ space
 \begin{eqnarray}
 G^{\phi\phi} (E,p)=G^{\phi\tilde\phi}(E,p)=G(E,p) ,
\nonumber\\
   G^{\tilde\phi\tilde\phi}(E,p)
=G(E,p)-G(E,\frac{1}{2})
\label{prpp}
\end{eqnarray}
where
\begin{eqnarray}
G (E,p)= { \sinh \pi E \over\pi  E  } {1\over   \cosh \pi E - \cos \pi p }
\nonumber \\
=  {2\over\pi^2}
  \sum _{n=-\infty}^{\infty}
{1 \over E^{2}+(p+2n)^{2}}.
\label{bleve}
\end{eqnarray}
The three propagators have the following diagrammatic meaning:
$G^{\tilde\phi\phi} $  ($ G^{\tilde\phi\tilde\phi}$)
 is  associated with  the external (internal) lines of a Feynman diagram,
and
 $G^{\phi\phi}$ is
 the genus-zero loop-loop correlator.

\subsection{Interactions}

 In order to compute the interaction potential we need the explicit answer
 for  the one-matrix integral.

The   free energy $F[V]$ of the
one-matrix model (\ref{fren}) is most easily expressed
not in terms of the potential $V(z)$, but in terms of the  field  $ \p\Phi(z) =
dF^{(0)}/dV(z) -\frac{1}{2}\partial
V(z)$ (\cite{dvv} - \cite{wi}).
The meromorphic function  $\p\Phi(z)$ has  vanishing real part along its
cut $[a=-\infty, b]$
and therefore can be expanded as a series in positive half-integer powers
of $z-b$. (Note that $b=b[V]$  depends on the potential!)
If $z_0$ is a point close to $b$, the field can be nevertheless
expanded as a series in the positive and negative   half-integer powers of
$z-z_0$.
 The free energy $F$ can be expressed in terms of the coefficients
of the nonsingular   at
$z=z_0$ part of the expansion.
The coefficients are given
 (up to numerical factors) by the linear functionals $ ( k|  \Phi)_{z_0},
k=0,1,...$  defined by the   generating function
\begin{eqnarray}
\sum_{k=0}^{\infty} {u^k\over k!} ( k|  \Phi)_{z_0} =
 \sqrt{2}  \oint {dz\over 2\pi i} {d  \Phi^c(z) / dz  \over  \sqrt{ z-z_0
-2u}}.
\end{eqnarray}
 Sometimes they are called KdV coordinates
of the field $\Phi$.
 The KdV coordinates at different expansion points are related by
\begin{eqnarray}
 (\Phi|n)_{z_0+2u}=   \sum _{k\ge 0} (\Phi|n+k)_{z_0} {u^k\over k!}.
 \label{hcub}
\end{eqnarray}

The choice of the expansion point $z_0$ is a matter of convenience. The
most compact
expression corresponds to the choice $z_0= b$, where $b$ is the position
of the branchpoint of
$\Phi(z)$,  imposed by the condition
\begin{eqnarray}
 (\Phi|0)_{b}= 0.
 \label{hbub}
\end{eqnarray}
In this case the  explicit expression for     of $F[\Phi] $
    reads  \cite{iz}
\begin{eqnarray}
F[\Phi ]  = F^{(0)}[\Phi ] -{1\over 24} \ln( 1 |\Phi)_b\nonumber \\
+\sum _{2g-2+n>0}
 ( 1 |\Phi)_b^{2-2g}
  {(-1)^n \over n!} \sum_{^{k_1,\ldots,k_n \ge 2}_{ k_1+\cdots+k_n=
3g-3+n}}
\nonumber\\
 \{  k_1 \cdots k_n \} _g\    {( k_1|  \Phi)_b\over
 (1|  \Phi)_b } \cdots
 {(k_n| \Phi)_b\over
  (1|  \Phi)_b}. \ \ \ \ \ \ \ \ \ \ \ \
\label{frpp}
\end{eqnarray}
 The coefficients $ \{ k_1\cdots k_n\}_g \ \ ( k_1+\cdots+k_n=
3g-3+n)$ are the
the intersection numbers on the moduli
space $\CM_{g,n}$ of algebraic curves  of genus $g$ with $n$ marked points
\cite{wi}.
 The intersection numbers can be
obtained from a system of recurrence relations equivalent to the loop
equations \cite{dvv}.
  In particular, the genus-zero the intersection numbers coincide with the
multinomial  coefficients
\begin{eqnarray}
 \{  k_1 \cdots k_n \} _0 =   {(k_1+\cdots+k_n)! \over k_1!\cdots k_n!} ,
\nonumber \\
k_1+\cdots+k_n= n-3.
\label{haha}
 \end{eqnarray}
  The genus $g$ term in the expansion of the free energy is given by the
restricted sum (\ref{frpp}) with  $ k_1+\cdots+k_n= 3g-3+n$, which
contains only finite number of terms.

 The expression of the free energy as a function of  the KdV coordinates
$(n|\Phi)_{z_0}$ with  fixed   $z_0 $  has the same form
but the sum over $k_1,...,k_n$ should be taken in the range $k_i\ge 0$
since the coordinate $(0|\Phi)_{z_0}$ no longer vanishes.
Note that $F$ is singular at $\Phi=0$ and has to be expanded around
some nontrivial background
 $ \Phi^c(z)$.
  The simplest nontrivial background to expand about is the
"topological" background
\begin{equation}
 \Phi^c(z)= {1\over \sqrt{2}} \frac{2}{3} \  {(z-z_0)^{3/2}  \over \kappa}
\label{topb}
\end{equation}
  with only
one nonvanishing coordinate $( \Phi^c|1)_{z_0}
 =1/\kappa $.   A generic potential can be considered as a deformation
about this topological point.
 For   deformation
parameters   $t_{k} =t_k(\Phi) $ defined as
 \begin{eqnarray}
(k|  \Phi)_{z_0}={1\over \kappa} (\delta_{k,1} - t_{k})
\end{eqnarray}
 the genus-$g$ term of the free energy reads, in terms of $t_{k}$,
   \begin{eqnarray}
  F^{(g)} [\Phi] =  \kappa ^{2g-2+n}
 \sum_{n\ge 0} {1\over n!}  \ \ \ \sum_{^{k_1,\ldots,k_n \ge 0}_{
k_1+\cdots+k_n= 3g-3+n}}
 \nonumber\\
\{  k_1 \cdots k_n \} _g\  t_{k_{1}}\cdots   t_{k_{n}}.
\end{eqnarray}
\smallskip

Now we are ready to find the  vertices (\ref{stvtx}) for the
 genus $g$, $n$-loop interactions ($2g+n-2>0$). We have  to expand the
effective potential
$F[\tilde \Phi_x]$ around the background (\ref{reshe}) and taking
$z_0=-1$.
Alternatively, we can expand around the topological background
(\ref{topb}) with $z_0=-1$
and consider the  background (\ref{reshe}) as a perturbation.
 We prefer to do the latter because in this case the
interaction vertices do not depend on the background.
Then the
 deformation
parameters   $t_{kx} =t_k(\tilde\phi_x) $  are  defined by
\begin{equation}
\delta_{k,1} - t_{kx}= {(k|\tilde \Phi_x) \over (1|\Phi^c_x)}.
\label{ttkj}
\end{equation}
(Once the expansion point is  $z_0=-1$ is fixed, we  will  write  $ (k|*  ) $
instead of
$  (k  |*)_{-1}$.)
We  know the explicit form of the propagator
and the tadpole in the base of plane waves, while the interaction is
expressed in terms of the
KdV coordinates.
Therefore  we need the
transformation matrix between these two mode expansions.
The generating function for the KdV coordinates is given, in the $\tau$
space, by
\begin{eqnarray}
\sum_{k=0}^{\infty} {u^k\over k!} (k|  \tilde\Phi) =
\int_0^{\infty}d\tau
    {\ptau  \cos \pi \ptau   \tilde\phi(\tau) \over   \sqrt{\cosh^2{\tau
\over 2} -u}}.
\label{ttaau}
\end{eqnarray}
 For  $ \tilde \phi (\tau) =  \sin E\tau$
the integral  is a Legendre function and its expansion in $u$ gives for
the
 KdV coordinates of a plane wave
\begin{equation}
 (k | E)=   \ \pi  E  \ \Pi_k(iE) , \ k=0, 1,2, ...
\label{pipi}
\end{equation}
where
\begin{eqnarray}
 \Pi_0(a)=1;
\nonumber \\
\Pi_k(a)= {1\over k!} \prod_{j=0}^{k-1}[(j+\frac{1}{2}   )^2 -a^2], \
k=1,2, ...
\label{pii}
\end{eqnarray}
Eq. (\ref{pipi}) means that the linear functional $( k| $ acts in the
space of  odd
functions  $\tilde\phi(\tau)= -\tilde\phi(-\tau)$   smooth   at
$\tau =0$ ,  as
 \begin{equation}
  ( k|\tilde\Phi)  = \pi   [\Pi_k (\ptau) \ \ptau \tilde\phi(\tau)
]_{\tau =0}.
\label{pipii }
\end{equation}

The KdV coordinates  of   the string background (\ref{reshe}) are obtained
using the
 identity  $ (  k | d\tilde\phi / d z  ) = \frac{1}{2}  \
(k+1|\tilde\phi)$
  \begin{equation}
(k|\tilde\phi ^c_x) =  -  {S_x\over\kappa}\ \Pi_{k-1} ( \beta ), \ k=2,3,
...  .
\label{teka}
\end{equation}

 The interaction vertices for the field $\tilde \Phi_x$ are obtained by
expanding  the free energy of the one-matrix model  around the
 deformation parameters  $t^c_k$ characterizing  the  saddle-point
solution. From (\ref{teka})
 we find
\begin{equation}
   t^c_0 = t^c_1 =0, \ \
   t^c_k =  -  \Pi_{k-1} ( \beta ), \ k=2,3, ...
\label{bertr})
\end{equation}

After going to the momentum space
the factor $S_x^{2g-2+n}$ in the genus expansion of $F[\Phi_x]$  yields
the  genus $g$ fusion coefficient (\ref{fcns})
    and  the  vertex factorises into a product of an intersection number
and a fusion coefficient. The KdV representation of the internal
propagator is   given by
 the symmetric matrix
\begin{eqnarray}
 G_{kk'}(p)= (k|(k'| G_{\tilde\phi\tilde\phi} (p) )); \\
   G_{00}(p)= -\Pi _1(1-p), \nonumber \\
  G_{01}(p)=G_{10}(p)=
-\Pi _2(1-p) ,\nonumber\\
 G_{11}(p)=-2\Pi_2(1-p)-2\Pi_3(1-p), ...
\label{intp}
\end{eqnarray}

Finally,
the  external line factors   read
   \begin{eqnarray}
(k , p| G_{\tilde\phi\phi}|\tau , p \rangle  =   \Pi_k({\p_\tau} )
{\sinh(1-p)\tau \over\sin \pi p\sinh\tau}.\label{exfc}
\end{eqnarray}

 To summarize, we have
found the following  loop Feynman rules:
\noindent

$Propagator:$
 \begin{eqnarray}
G_{k k'}(p), \ \ p={m\over h}
\end{eqnarray}

$Tadpole:$
 \begin{eqnarray}
V^{(0)}_{k,m} = {1\over \kappa} t^c_{k} \delta_{m,1}
\end{eqnarray}

 $Vertices: \ \ \ \ (g+2n\ge 2)$
   \begin{eqnarray}
  V^{(g)}_{k_1,m_1,...,k_n,m_n} \nonumber \\
 =  \kappa ^{2g-2+n}
  \{  k_1 \cdots k_n \} _g\  C^{(g)}_{m_1...m_n}.
\label{oijkn}
\end{eqnarray}
Let us  give several examples.

\noindent $(i)$  The   three-loop genus-zero amplitude
\begin{eqnarray}
  \langle    \Phi_{m_1} (z_1) \Phi_{m_2} (z_2)  \Phi_{m_3} (z_3)\rangle_0=
\nonumber \\
\kappa  C_{m_1m_2m_3}^0 \prod _{s=1}^3
   {\sinh ({h-m_s\over h})\tau_s \over \sin \pi {m_s\over h} \sinh
\tau_s}.\label{thrt}
\end{eqnarray}

\noindent $(ii)$  The genus-one one-loop amplitude
 \begin{eqnarray}
  \langle   \Phi_{m} (z) \rangle_1 =  \kappa
\Big[  C^1_{m} \{ 1 \}_1  \Pi_1(\ptau)
+   C^1_{m} \{ 02\}_1 t_2^c \nonumber \\
+  {1\over 2}  \sum_{m'} C^0_{mm'm'} \{000\}_0
 G_{00}(m')  \Big]  {\sinh ({h-m\over m})\tau \over \sin \pi {m\over h}
\sinh \tau} \nonumber \\
 = \kappa \delta_{m,1}
\sum _{m'}\Big[{1\over 24}   \Big(\beta^2 - {\p^2 \over \p \tau^2}\Big)  +
{1\over 2}\Big({m'\over h} -{1\over 2}\Big)\nonumber \\
\Big({m'\over h}-{3\over 2}\Big) \Big]
{\sinh ({h-1\over h})\tau \over \sin {\pi\over h}  \sinh \tau}. \ \ \ \ \
\ \ \ \ \ \ \ \ \ \ \ \ \ \ \ \ \ \ \ \ \ \
\label{oonnl}
 \end{eqnarray}

\noindent $(iii)$ The four-loop genus-zero amplitude
  \begin{eqnarray}
\langle \prod _{s=1}^4    \Phi_{m_s}(z_s)\rangle_0=
\kappa^2  \Bigg[\Big[   \beta ^2-{1\over 4}
  +  \sum_{s=1}^4 \Big({1\over 4 }\ \ \ \ \ \nonumber \\-
{\p^2\over\p\tau_s^2}\Big)
 \Big]C^0_{m_1m_2m_3m_4}  + \sum_m  \Big( C^0_{m_1m_2m}
C^0_{m m_3m_4}\nonumber \\ +C^0_{m_1m_3m}  C^0_{m
m_2m_4}+C^0_{m_1m_4m}C^0_{m m_2m_3}\Big)\ \ \ \
 \nonumber \\
  \Big({m\over h}-{1\over 2}\Big)\Big({m\over h}-{3\over 2}\Big)  \Bigg]
\ \ \prod_{s=1}^4  {\sinh (1-{m_s\over h})\tau_s \over \sin \pi {m_s\over
h} \sinh \tau_s}.
\nonumber \\
\label{iyt}
 \end{eqnarray}
 The expression for the genus 1 tadpole (\ref{oonnl}) is in accord with
the recent calculation  by B. Eynard and C. Kristjansen in the $O(n)$
model. (In this case
one has to put $\pi m/h =   \arccos (n/2).$ )
Eq. (\ref{iyt}) reproduces in the limit $R\to\infty$ the four-loop
amplitude found
in ref. \cite{koSt}.

\section{ VERTEX OPERATOR CONSTRUCTION}

 The relation between the Coulomb gas systems and the integrable hierarchies of
differential soliton equations gives another powerful approach to study these
systems, especially in application to nonperturbative phenomena.

  The relevance of the integrable hierarchies to the critical phenomena on
random surfaces is now an established fact.
It has been shown by M. Douglas   \cite{Mike} that the chains of random
matrices  with nearest neighbor interaction
   ($\tr {\bf M_x}{\bf M_{x+1}}$)  are described by the $A$ series of
generalized KdV flows in the
classification of Drinfeld and Sokolov \cite{DS}. Moreover, as it was noticed
by
Di Francesco and Kutasov \cite{DFK},   a particular case of the
$D$ series   reproduces the known qualitative features of the $D$  models
of matter fields coupled to   gravity .

It remains an open problem, however, how to construct the hierarchies
associated with the  rational conformal theories coupled to gravity in a
unified fashion, including the exceptional ones.
We believe  that the $ADE$ and $\hat A\hat D\hat E$ matrix models
can be used as a starting point to solve this problem.
The logic is the following. Each matrix model partition function
in the form of  Coulomb gas defines a bosonic vertex operator construction.
Then the associated hierarchy of soliton equations can be written
as a system of Hirota bilinear equations \cite{hir} following the general
prescription proposed in \cite{KW}. In the case $\hat A_0$ this is   KdV
hierarchy
 (just as in the case of the $A_1$  model, but of course
  with different boundary conditions).

 Below  we show how to make  the first step, namely the bosonic vertex operator
construction.  We restrict ourselves to  the $A$ and $\hat A$ series, but the
generalization to the other cases is evident.
   Then we  give the fermionic representation for the  $A$ and $\hat A$ series.
The boson-fermion correspondence allows to
establish  a fermionic representation  for the    $A, \hat A, D$ and $\hat D$
series, but not for the
  exceptional ($E$ and $\hat E$) cases.
There should be certainly a connection
 between the  hierarchies associated with the $A$ and $D$
Coulomb systems and   the  Lax representations
by Drinfeld and Sokolov  for the $A$ and $D$ series \cite{DS}.

   \subsection{The case $\hat A_0$}

  Before presenting the general construction let us consider
   in detail the simplest nontrivial example, namely the  $\hat A_0$
   model  whose the target space is a
circle with one point.

Let us first note that if the integration with respect to $\lambda$ is
replaced by contour integration with respect to the complex variables $z_i$,
the partition function will still satisfy the same loop equations
(\ref{vviir}).

We start with the construction of  the bosonic field representation of the
canonical
 partition function
   \begin{eqnarray}\CZ_{\hat A_0} = \sum_{N=0}^{\infty}
 \int \prod _{i=1}^N dz_i \ e^{-V(z_i)}  {\prod_{i< j} (z_i-z_j)^2\over
  \prod_{ij}(-z_i-z_j)}.
   \label{odve}
\end{eqnarray}

 Introduce  the  bosonic field
$\varphi(z) $ with mode expansion
   \begin{eqnarray}
  \varphi(z) = \hat q+ \hat p\ln z +\sum_{n\ne 0}  {J_{n}\over n}
z^{-n},
\end{eqnarray}
\begin{eqnarray}
[ J_n,  J_m]
=n\delta_{m+n, 0};   \ \  [\hat p,\hat q] = 1.
\label{henno}
\end{eqnarray}
and the vacuum state $|l\rangle$  defined by
\begin{equation}
J_n|l\rangle =0 , \ \  (n>0); \ \ \  \hat p |l\rangle=l|l\rangle .
\end{equation}
The associated normal ordering is defined by putting $J_n, n>0$ to the right.
 Following ref. \cite{LeWi} we  define the antisymmetric field
  \begin{eqnarray}
 \Phi(z) = \varphi (z)
 - \varphi(-z).
\label{szso}
\end{eqnarray}
 The field $\Phi(z)$ and the vertex operator $:e^{\Phi(z)}:$
satisfy
  the OPE
 \begin{eqnarray}
\Phi (z)\Phi (z')= :\Phi (z)\Phi (z'):
\nonumber \\
+2 \ln (z-z') -2\ln (z+z'),
\label{Glfo}
\end{eqnarray}
  \begin{eqnarray} {:e^{\Phi(z)}: :e^{\Phi(z')}:
  = {(z-z') ^{2} \over (z+z')
^{2} }  }
:e^{\Phi(z)}e^{\Phi(z')}:.
  \label{opevo}
\end{eqnarray}
Define the Hamiltonian
\begin{equation}
H[t]=\sum_{n}t_{n}J_n,
\label{hahaha}
\end{equation}
and the operator
\begin{equation}
G_{A_0} = \exp \Big( e^{\mu} \oint {dz\over 2\pi i}  :e^{ \Phi(z)}: \Big).
\label{gegege}
\end{equation}
 It is easy to see that the vacuum expectation value
  \begin{eqnarray}\tau_0[t]  = \langle 0|
 e^{ H[t]} G _{A_0}|0\rangle
\label{pYiao}
\end{eqnarray}
is equal to the partition function (\ref{odve})
 with potential
\begin{equation}
V(z)= -\mu - \sum_{n\ge 0}t_{2n+1}z^{2n+1}.
\label{plopl}
\end{equation}

The fermionic representation of the partition function of the $\hat A_0$
model follows from the bosonization formulas
  \begin{eqnarray}\psi(z)= :e^{-\varphi(z)}: \ ,
 \psi^*(z)=
:e^{\varphi(z)}: \nonumber \\
  \partial\varphi(z)= :\psi^*(z)
\psi(z):
 \label{bBo}
\end{eqnarray}
 where the
  fermion operators
 \begin{eqnarray}
	\psi(z)= \sum_{r\in \Z+\frac{1}{2}}\psi_{r}
z^{-r-\frac{1}{2}}\nonumber \\
	\psi^*(z)= \sum_{r\in \Z+\frac{1}{2}}\psi^*_{-r} z^{-r-\frac{1}{2}}
	\label{pzpo}
\end{eqnarray}
satisfy the modes in the expansion of the anticommutation relations
 \begin{eqnarray}
 	[\psi_{r},\psi^*_{s}]_+=\delta_{rs}.
 	  \label{cpmto}
 \end{eqnarray}
The  operators (\ref{hahaha}) and (\ref{gegege}) are represented by
 \begin{equation}
 	H[t] = \sum  _{n>0}t_n  \sum_r :\psi^*_{r-n}\psi_{r}: \ \ (n\in \Z)
 	\label{curro}
 \end{equation}
 \begin{eqnarray}
	G=\exp\Big[ e^{\mu} \oint {dz\over 2\pi i}
:\psi(z)\psi^*(-z) :
	\Big].
	\label{araro}
\end{eqnarray}
 and the
vacuum states with given electric charge $l$ satisfy
 \begin{eqnarray}
	\langle  l|	\psi_{-r} =\langle l| \psi^*_{r}= 0\ \
	 \ (r>l )\\
\psi_{r}| l \rangle =\psi^*_{-r}|l\rangle = 0\ \
 \ (r> l).
	\label{mnfio}
\end{eqnarray}

Following the line of arguments of ref. \cite{babe}, we can consider
  the  canonical partition function  (\ref{odve})  as the large $N$
    limit of  $N$-soliton solutions of the KdV hierarchy of
integrable  differential equations (where $t_1$ is
     the space variable and $t_3, t_5,.. $ the time variables).
      Therefore the partition function  (\ref{odve})
 represents itself a $\tau$-function
      of the KdV hierarchy.

More generally, one can define
  \begin{eqnarray}\tau_l[t]  = \langle l|
 e^{ H[t]} G _{A_0}|l\rangle .
\label{pYiall}
\end{eqnarray}
 for any integer $l$ but  in fact  there are only two different
 $\tau$-functions corresponding to $l=0 (mod\ 2)$ and $l=1 (mod\ 2)$.
   In terms of Fredholm determinants
\begin{equation}
\tau_{l}= \det (1+ (-)^l K)
\label{frtt}
\end{equation}
where the kernel $K$ is defined by (\ref{kernoo}).

The KdV and mKdV hierarchies of differential equations
  are generated by the Hirota bilinear equations \cite{hir}
\begin{eqnarray}
\oint{dz\over 2\pi i} z^{l-l'}
\exp\Big(\sum_n(t_n-t'_n) z^n\Big) \nonumber \\
\tau_{l}(t_n-{1\over n} z^{-n})
\ \tau_{l'}(t'_n+{1\over n} z^{-n}) = 0 .
\label{Hiir}
\end{eqnarray}
Namely, the second derivative
      \begin{equation}
      u=2 {\p^2\over \p t_1^2} \log \tau_0
      \end{equation}
      satisfies the KdV hierarchy of differential equations, the
      first of which is the classical KdV equation
      \begin{equation}
      {\p u\over \p t_3}=6 u{\p u\over \p t_1}+{\p^3
      u\over \p t_1^3}.
      \label{KDV}
      \end{equation}
       Further, the function
         \begin{equation}
         v= {\p\over \p t_1} \log{\tau_0\over \tau_1}
        \end{equation}
        satisfies the modified KdV equation
        \begin{equation}
          {\p v\over \p t_3}=-6 v^2{\p v\over \p t_1}+ {\p^3 v\over \p t_1^3}
      \label{mKDV}
      \end{equation}
      and $u$ and $v$ are related by the Miura transformation
      \begin{equation}
      u=-v^2-{\p v\over t_1}.
      \end{equation}

Finally, the bosonic and fermionic
representations generalize trivially   if we add negative odd powers to the
potential,
   \begin{equation}
   \tau_{l} [t] = \langle l|e^{\sum_{n\ge 0}t_nz^n } \ G_{A_0}\ e^{\sum_{n<
0}t_nz^n }|l\rangle,
   \label{ttatra}
   \end{equation}
   and the $\tau$-functions (\ref{ttatra}) solve  the affine sinh-Gordon
hierarchy
   (with $t_{-1}, t_{-3},... $
   as time variables). The lowest order equations are
   \begin{eqnarray}
  { \p^2 \phi\over \p t_1\p t_{-1}}={1\over 2} \sinh 2\phi , \\
  -{ \p^2  \over \p t_1\p t_{-1}}\log \tau_1 ={e^{2\phi}-1\over 4}
\\
(\phi = \log {\tau_0\over \tau_1}).\nonumber
  \label{shgr}
  \end{eqnarray}
 The relation between the Fredholm determinants (\ref{frtt}) and
 the mKdV and sine-Gordon hierarchy
has been established directly  in  \cite{trw}.

\subsection{  Bosonic construction in the general case   }

The idea is to construct the
operators $\Phi ^x(z), \ x\in X, $  associated with the roots of the
corresponding
classical Lie algebra,
with OPE
  \begin{eqnarray} \Phi ^x(z)\Phi ^{x'}(z')= :\Phi ^x(z)\Phi ^{x'}(z'):
\nonumber \\
+2\delta_{xx'} \ln (z-z') -A^{xx'}\ln (z+z').
\label{Glf}
\end{eqnarray}
 We will restrict ourselves to the cases $A_r$ and $\hat A_r$;
the other simply laced algebras can be considered in a similar way.

 We will use the fact that the  simple roots
$ \vec \alpha^x \ (x=1,...,r )$ and the lowest root $\vec \alpha^0$ of $A_{r}$
 can be represented as differences
of orthonormal vectors $\vec e_{1},...,\vec e_{r+1} $  in
$\R^{r+1}$
\begin{eqnarray}
 \vec \alpha^x= \vec e_x - \vec e_{x+1} ,  \ \ \
(x=0,...,r ) \nonumber\\
\vec \alpha^0= \vec e_{r+1} - \vec e_{1}. \ \ \ \ \ \ \ \ \ \ \ \ \ \ \ \ \ \ \
\
 \label{arara}
  \end{eqnarray}
  With each vector  $\vec e_a  \ ( a= 1, 2, ..., r+1) $
  we associate  the  bosonic field
$\varphi_a(z)\ \ $
  having mode expansion
   \begin{eqnarray}
  \varphi_a(z) = \hat q_a+ \hat p_a\ln z +\sum_{n\ne 0}  {J_{an}\over n}
z^{-n},
  \label{skf}
\end{eqnarray}
with
\begin{eqnarray}
 [J_{an} ,J_{a'm} ]= n \delta_{a,a'} \delta_{n+m, 0},
\end{eqnarray}
\begin{eqnarray}
  [\hat p_a,\hat q_{a'}] = \delta_{a,a'}.
\label{henn}
\end{eqnarray}
The bosonic Fock space $\CF_{\vec l } , \ \vec l \equiv \{ l_1,...,l_{r+1}\}$
is
generated by the action of the negative modes on the vacuum state $|\vec
l\rangle = e^{\sum   l _a   \hat q _a }|\vec 0\rangle $ satisfying
\begin{eqnarray}
 \hat p _a |\vec l\rangle= l_a|\vec l\rangle; \ \ \
 J_{an} |\vec l\rangle =0,   \ n>0.
 \label{vvva}
\end{eqnarray}
The state $\langle \vec l |$ is defined similarly, with the normalization
$\langle \vec l |
\vec l'\rangle= \prod_a \delta_{l_a, l'_a}$.

 The   fields
 \begin{eqnarray}
 \Phi^x (z) =  \varphi_x(z) - \varphi_{x+1}(-z) \ \ (x=1,...,r)
\nonumber\\
  \Phi^0 (z) =  \varphi_{r+1}(z) - \varphi_{1}(-z)
\label{pphhii}
\end{eqnarray}
satisfy (\ref{Glf}) and therefore the corresponding
 vertex operators  have  OPE
  \begin{eqnarray} {:e^{\Phi^x(z)}: :e^{\Phi^{x'}(z')}:
  = {(z-z') ^{2\delta _{xx'}} \over (z+z')
^{\delta_{x,x'+1}+\delta_{x,x'-1}} }  }\nonumber \\ \ \ \ \ \ \ \ \ \ \ \ \
:e^{\Phi^x(z)}e^{\Phi^x(z')}:.
  \label{opev}
\end{eqnarray}

Define the  Hamiltonian
  \begin{eqnarray}   H[\vec t\ ]=\sum _{a
=1} ^{r+1}
\sum_{n=0}^{\infty} t_{an} J_{an} .
\label{cohs}
\end{eqnarray}
and the operator creating the electric charges with fugacity $e^{\mu}$
\begin{equation}
	G_{\hat A_r} = \exp\Bigg(e^{\mu} \sum_{x=0}^r:  \oint {dz\over 2\pi i}
e^{\Phi^x(z)}:\Bigg).
	\label{glin}
\end{equation}
Then the vacuum expectation value
\begin{eqnarray}
 \tau_{\vec 0}[\vec t, \mu ] = \langle  \vec  0 | e^{  H [\vec t\ ]}  G |\vec
0\rangle
\label{ptasu}
\end{eqnarray}
is equal to  the canonical  partition function  (\ref{mtria})
for the model $\hat A_r$
  with
potential
  \begin{eqnarray} V^x(z) = -\mu  -\sum_n z^n [t_{x n}   -( -)^n t_ {x+1,n}].
\label{ppoo}
\end{eqnarray}

More generally, one can define the $\tau$-functions associated with
vacuum states with different vector charges $\vec l$ and $\vec l'$
such that
 $l_1 +...+l_{r+1}=  l_1' +...+  l_{r+1}'$
\begin{eqnarray}
 \tau_{\vec l , \vec \l' }[\vec t\ ] = \langle  \vec   l | e^{  H[\vec t\ ] }
 G _{\hat A_r}|\vec l'\rangle.
\label{tgen}
\end{eqnarray}

  The bosonic representation   of the $A_r$ model
 is constructed in the same way.
The canonical partition function is
  given by
\begin{eqnarray} \CZ _{ A_r}[\vec t]=
\langle \vec 0 | e^{  H [\vec t\ ]}  G _{A_r}|\vec 0\rangle
\label{ptria}
\end{eqnarray}
where
 \begin{equation}
	G_{A_r}= \exp\Bigg(  \sum_{x=1}^r  \oint {dz\over 2\pi i} :e^{
\Phi^x(z)}:\Bigg)
	\label{glino}
\end{equation}
and
\begin{eqnarray}
t_{x0}-t_{x+1, 0} \equiv \mu^x = {S^x\over \kappa }
\nonumber \\
S^x=2S_x-A_{xx'}S_{x'}= 4\sin^2 \Big({\pi\over 2h}\Big)\  S_x.
 \end{eqnarray}

\subsection{Fermionic representation of the $\hat A_r$ and $A_r$ models with
$r\ge 2$}

  The fermionic representation of the partition functions   follows from the
bosonization formulas
 \begin{eqnarray}\psi_a(z)= :e^{-\varphi_a(z)}:, \\
 \psi_a^*(z)=
:e^{\varphi_a(z)}:, \\
  \partial\varphi_a(z)= :\psi^*_a(z)
\psi_a(z):.
 \label{bB}
\end{eqnarray}

Let us introduce the fermion operators $\psi_{r,a},\psi^*_{r,a}\ (r\in
\Z+\frac{1}{2},
a=1,  ...,R+1)$ satisfying the
anticommutation relations
 \begin{eqnarray}
 	[\psi_{r,a},\psi^*_{s,b}]_+=\delta_{rs}\delta_{ab} .
 	  \label{cpmt}
 \end{eqnarray}
Define the zero-particle states $|-\infty\rangle, \langle -\infty|$
by
\begin{eqnarray}
\psi_{r,a}|-\infty\rangle =0,
\langle -\infty| \psi^*_{r,a}=0
\end{eqnarray}
and the  states  $|\vec l\rangle = |  l_{1},...,l_{r+1}\rangle ,\ \
\langle \vec l| = \langle  l_{1},...,l_{r+1}|, \ \ (\l_a\in \Z)$ by
 \begin{eqnarray}
|\vec l \rangle =\prod_a \prod_{r<l_a}\psi^*_{r,a}|-\infty\rangle
	\nonumber \\
	\langle\vec l | =\langle -\infty|\prod_a \prod_{r<l_a}\psi_{r,a}.
 \end{eqnarray}
These states satisfy
\begin{eqnarray}
	\langle \vec l|	\psi_{-r,a} =\langle \vec l| \psi^*_{r,a}= 0\ \
	 \ (r>l_a )\\
\psi_{r,a}|\vec l \rangle =\psi^*_{-r,a}|\vec l\rangle = 0\ \
 \ (r> l_a) .
	\label{mnfi}
\end{eqnarray}

Define the current operators
 $J_{n,a}$
 \begin{equation}
 	J_{n,a}= \sum_r :\psi^*_{r-n,a}\psi_{r,a}: \ \ (n\in \Z)
 	\label{curr}
 \end{equation}
 where
  \begin{equation}
	:\psi_{r,a}^*\psi_{s,b}: = \psi_{r,a}^*\psi_{sb}
-\langle 0|\psi_{r,a}^*\psi_{sb}
|0\rangle ,
	\label{nror}
\end{equation}
 or, equivalently,
 \begin{equation}
J_a(z)=\sum_{n\in \Z}J_{n,a} \ z^{-n-1} = :\psi_a^* (z)\psi_a(z):
 	\label{jzjz}
 \end{equation}
where
\begin{eqnarray}
	\psi_a(z)= \sum_{r\in \Z+\frac{1}{2}}\psi_{r,a}
z^{-r-\frac{1}{2}},\nonumber \\
	\psi^*_a(z)= \sum_{r\in \Z+\frac{1}{2}}\psi^*_{-r,a} z^{-r-\frac{1}{2}}.
	\label{pzpz}
\end{eqnarray}
Introduce the Hamiltonian  $H$ by (\ref{cohs})
  such that
\begin{eqnarray}
e^{H(t)} \psi_a(z)e^{-H(t)}= e^{\sum_{n=1}^{\infty}  t_{n,a}z^n}
\psi_a(z)
\nonumber \\
e^{H(t)} \psi_a^*(z)e^{-H(t)}= e^{-
\sum_{n=1}^{\infty}  t_{n,a}z^n}
 \psi_a^*(z).
\label{hhhps}
\end{eqnarray}
Then the operators (\ref{glin}) and (\ref{glino})  in the definition of the
$\tau$ functions  (\ref{ptasu}) and (\ref{ptria})
 are represented by
 \begin{eqnarray}
	G_{\hat A_r}=\exp e^{\mu}  \oint {dz\over 2\pi i} \Big[ \sum _{x=1}^{r}
:\psi_x(z)\psi^*_{x+1}(-z):
	\nonumber \\
	 +:\psi_{r+1}(z)\psi^*_{1}(-z):\Big] .
	\label{araf}
	\end{eqnarray}
and
 \begin{eqnarray}
	G_{A_r} = \exp \Big[\oint {dz\over 2\pi i} \sum_{x=1}^{r}
:\psi_{x}(z)\psi^*_{x+1}(-z) :\Big]
 . 	\label{araraa}
\end{eqnarray}

The  $\tau$-functions for the $A_r$ and $\hat A_r$ models  satisfy the
Hirota
bilinear equations \cite{hir} following from the
commutatativity of the  operators $S=\sum_a\oint dz \psi_a(z)\otimes
\psi^*_a (z)$
 and  $G\otimes G$.
Bilinear equations for the $D$ and $E$ series can be written as well but
they have  more complicated
form \cite{KW}.

 \end{document}